# An exact model for predicting tablet and blend content uniformity based on the theory of fluctuations in mixtures

## Sagar S. Rane*, Ehab Hamed, and Sarah Rieschl

Cima Labs Inc.
R&D Formulations
7325 Aspen Lane
Brooklyn Park, MN 55428

**Abstract**

The content uniformity (CU) of blend and tablet formulations is a critical property that needs to be well controlled in order to produce an acceptable pharmaceutical product. Methods that allow the formulations scientist to predict the CU accurately can greatly help in reducing the development efforts. This article presents a new statistical mechanical framework for predicting CU based on first principles at the molecular level. The tablet is modeled as an open system which can be treated as a grand canonical ensemble to calculate fluctuations in the number of granules and thus the CU. Exact analytical solutions to hard sphere mixture systems available in the literature are applied to derive an expression for the CU and elucidate the different factors that impact CU. It is shown that there is a single ratio, $\lambda \equiv <w^2.f^2>/<w.f>$; that completely characterizes "granule quality" with respect to impact on CU. Here $w$ and $f$ denote the weight of granule and the fractional (w/w) assay of API in it. This ratio should be as small as possible to obtain best CU. We also derive analytical expressions which show how the granule loading impacts the CU through the excluded volume, which has been largely ignored in the literature to date. The model was tested against literature data and a large set of tablet formulations specifically made and analyzed for CU using a model API. The formulations covered the effect of granule size, percentage loading, and tablet weight on the CU. The model is able to predict the mean experimental coefficient of variation ($CV$) with good success and captures all the elements that impact the CU. The predictions of the model serve as a theoretical lower limit for the mean $CV$ (for infinite batches or tablets) that can be expected during manufacturing assuming the best processing conditions.

*Keywords*: content uniformity, formulation, tablet, tableting, mathematical model, powder technology, thermodynamics/statistical mechanics, solid dosage form, molecular modeling, particle size

**Table of contents graphic**

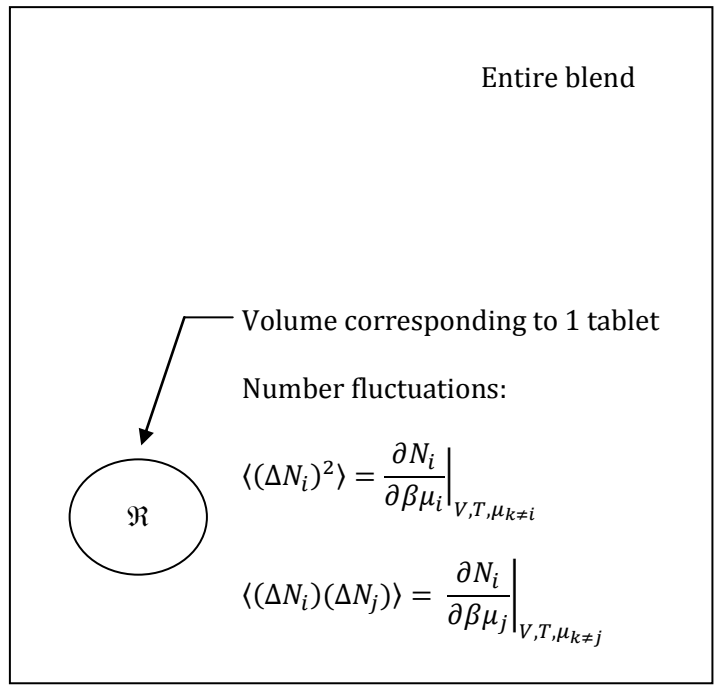

**INTRODUCTION**

Granulation is a commonly used route in pharmaceutical product development to increase dissolution rate, flowability, density of API, better distribution of API for low dose formulations, or in the case of multiparticulate systems to provide for a solid core for coating a release modifying polymer. However, when granules are added to a blend formulation, problems of poor content uniformity (CU) may arise due to a combination of factors such as non-ideal granule characteristics (particle size, assay distribution), low tablet dose, particle size differences with excipients, and segregation during manufacturing operations. Therefore there is great interest in the industry to understand the interplay of various granule characteristics such as assay variation with particle size, particle size distribution, dose, and granule loading with tablet CU. Such an understanding can help identify conditions that would yield the best CU in tablets. Here an exact theoretical model for predicting the CU assuming random mixing of granules but allowing for packing effects within granules is presented. The results are general in nature and are applicable to tablet formulations with particles of any size or type such as layered/coated beads,[1] pellets,[2] microspheres,[3] or others. Thus, the current model can have a broad impact on the field.

A number of publications have addressed the issue of predicting tablet content uniformity from particle size distribution of the API. Different theoretical approaches have been implemented. Johnson presented simple and elegant analytical calculations for the expected content uniformity for a given API particle size distribution assuming a Poisson distribution of API particles in the tablets.[4] The approach was extended by Yalkowsky et al[5] by assuming normal distribution of tablet assay. Yalkowsky et al. presented another analytical expression for the content uniformity from mean particle size and relative standard deviation. They also calculated an upper limit for mean particle diameter for a given relative standard deviation with log-normal particle size distribution which assured that the USP CU criteria was met with 99% probability. Rohrs et al.[6] took a step further and incorporated the effect of testing limited sample size using the chi-square distribution. They identified upper limits for the particle mean size in order to pass the USP CU criteria with 99% confidence for a given particle standard deviation and target dose. Recently, Huang et al.[7] presented a numerical simulation of tablet content uniformity from API particle size using the Monte Carlo simulation strategies. A nomograph was developed that can be used to define cut-off particle size requirements to pass the USP CU criteria with 99% certainty at a target dose. They also showed that the coefficient of variation and skewness are decoupled from the API particle size distribution and that the coefficient of variation varies inversely with the square root of the dose. Egermann et al have developed a model for tablet content uniformity (binary mixtures only) based on the binomial distribution and percolation theory.[8-10] Thus, a good amount of insight into how API particle size determines the content uniformity is available in the literature. However, there is no information in the literature for predicting CU in multiparticulate formulations that use granules or beads.

In addition, the impact of granule loading (as a proportion of tablet size) on content uniformity has not been systematically evaluated in the literature. Nor is it clear whether the observed relationships for tablets made with pure API particles would hold for tablets made with beads or granules. The impact of granule loading on the content uniformity has been discussed to some extent for e.g. by Johnson[4] and Egermann.[8,9] Assuming non-random mixing (i.e. particle packing effects) in a two component system, Johnson had proposed that when the API loading is between 1-10% w/w the coefficient of variation (*CV*) can be predicted roughly by multiplying the *CV* observed under ideal mixing conditions with the weight fraction of the excipient. As commented by Johnson, extending non-random mixing to multicomponent systems is complicated. Egermann's approach on the other hand is valid only for binary mixtures and does not take into account the particle size distribution of granules. Therefore, no general or first principle based solutions are available for incorporating the effect of the granule loading which, as we show herein, has a large impact on tablet CU regardless of the dose.

All of the work in the literature[4-10] thus far is based on the assumption of ideal mixing, with the exception of some results due to Johnson[4] and Egermann.[8,9] In ideal mixing, no regard is given to excluded volume which plays an important role in determining the radial distribution function and thus the content uniformity. Excluded volume refers to the actual volume occupied by the granules in a tablet, i.e. volume which is excluded for placement of other particles. Taking account of excluded volume ensures that no two particles occupy the same space. Ideal mixing is only truly valid when the excluded volume of API or granules is very small (for example at low doses). It is due to this reason that the current models in the literature cannot account for the effect of granule loading or tablet size on the content uniformity.

Statistical mechanical models[11-20] or computer simulations[21-24] can yield deep insight into the structure of materials at the molecular level. We show that the problem of predicting content uniformity in blend or tablet is essentially of the same category and can be solved with the framework of statistical mechanics.

In this article, analytical theory for hard sphere mixtures is implemented to calculate the number fluctuations in granules and hence the *exact CU under the assumption of random mixing while allowing for packing effects within granules*. Analytical solution to the problem of hard spheres became possible after the Percus-Yevick (P-Y) approximation[11,12] was used to solve the Ornstein-Zernike equation.[13] Wertheim[14] and Thiele[15] independently obtained an exact solution for the radial distribution function of a fluid of single component hard spheres using the P-Y approximation. While Lebowitz[16,17] obtained an exact solution for the fluid of a mixture of hard spheres. The equation of states obtained for both the single component and mixture of hard spheres have been shown to be in very good agreement with computer simulations for the whole range of fluid densities. Therefore, these exact analytical solutions provide an important foundation for calculating fluid properties and are implemented in the current work. By modeling the granules as hard spheres we can take into account the excluded volume and thus account for the effect of granule loading and packing on the *CV*. In the current approach, excipients are ignored because the results are obtained under the assumption of random mixing of granules. Random mixing is a commonly used approach in statistical mechanics to understand the behavior of complex systems and is a good approximation in situations where there is weak interaction between particles.

Since the current approach is based on equilibrium thermodynamics, it predicts the structure of the blend under equilibrium conditions only, without regard to the path taken to achieve that state. If one desires to know the path taken to achieve the final CU, other theoretical methods or simulations such as DEM (discrete element method) will be needed which calculate particle trajectories during processing operations.[25] Since the goal of every system is to achieve equilibrium conditions, the current approach yields valuable insight into molecular factors that govern the final CU.

Finally, to confirm the theoretical model, its predictions were compared with literature data as well as an extensive tablet design study. The experimental study was conducted using two lots of granules manufactured from a model API. Tablets with varying dose, weight, and granule loading were manufactured and analyzed for CU. Good agreement was obtained between the predictions of the theoretical model and the experimental *CV*. This shows that the model is able to capture all the elements that impact the *CV* under the assumptions of random mixing.

**MATERIALS**

Diphenhydramine HCL USP was obtained from Letco Medical, Inc (Decatur, AL) with a reported purity of >99%. The API had $d_{10}$, $d_{50}$, and $d_{90}$ of 21, 84, and 195 micron respectively. Mannitol 60 (Pearlitol® 160C) was obtained from Roquette Frères, France. Granular mannitol 2080 (Mannogem™) was obtained from SPI Pharma (Grand Haven, MI). Microcrystalline cellulose (Avicel®) was obtained from FMC Biopolymer (Philadelphia, PA). Crospovidone (Polyplasdone® XL) was obtained from ISP technologies (Calvert city, KY). Magnesium stearate was obtained from Mallinckrodt Inc (St. Louis, MO). Eudragit® E100 was obtained from Evonik Röhm Pharma Polymers, Germany. All of the above excipients complied with USP and Ph. Eur specifications and were used as obtained without any further processing. Ethyl alcohol USP was obtained from Grain Processing corporation (Muscatine, IA). Squalene (purity> 98%) was obtained from Sigma-Aldrich (St. Louis, MO).

**EXPERIMENTAL SECTION**

**Theoretical Model**

In this article, a new approach to predicting content uniformity of granules in tablets or blend is presented. The approach is based on modeling the blend as a fluid mixture and applying the theory of fluctuations derived from statistical mechanics. This approach, combined with analytical theory of hard sphere fluid mixtures, allows us to derive a *direct* and *exact* model for the content uniformity based on the assumption of random mixing. A scheme illustrating the approach is depicted in Figure 1.

Consider a volume in a given blend which corresponds to one tablet mass after accounting for the blend density (region ℜ in Figure 1). This blend volume can be considered to be ultimately compressed into a tablet and therefore studying the drug content in this volume allows us to obtain an expression for the content uniformity. Let *m* denote the total number of granule types or species in the blend. Here a granule type is defined as one which has a unique particle size and associated assay. (Note: Granule type can be defined using any arbitrary criteria and is not limited to particle size/assay. However, in the present problem defining the type using these criteria is most relevant as shown in the included examples) We assume that each granule is impenetrable and is spherical. Let $\mu_i$ denote the chemical potential for granules of diameter $R_i$.

Region ℜ in figure 1 is "open" i.e. it can allow for granules to enter and leave. Note that the volume *V* (corresponding to one tablet) of the region ℜ is fixed. The actual number of granules of each type *i*, $N_i$, in region ℜ is controlled by the respective chemical potential, $\mu_i$. In statistical mechanics, such a system where the volume, temperature, and chemical potentials are fixed while allowing for the number of granules or particles to fluctuate is referred to as the grand canonical ensemble.[23]

Let $w_i$ and $f_i$ denote the weight of each granule of type $i$ and the fractional (w/w) assay of API in it. Then, the weight of the API in each granule of type $i$ is $a_i = w_i.f_i$. Each tablet has a variable number of total number of granules ($N$) and there is a distribution in the weight of the granules. Let $\bar{N}_i$ denote the average number of granules of type $i$ in region $\mathfrak{R}$. The variance in the number of granules of each type $i$ is given as:

$$\text{Var}(N_i) = \langle (N_i - \bar{N}_i)^2 \rangle = \langle (\Delta N_i)^2 \rangle \tag{1}$$

Thus $\text{Var}(N_i)$ is identical to the mean square number fluctuations of granules of type $i$.

The dose is given as,

$$D = \langle \sum_{j=1}^{N} \hat{a}_j \rangle = <N><w.f> \tag{2}$$

where $\hat{a}_j$ is the weight of the API in granule $j$. The variance in the dose can be written as,[5]

$$\begin{aligned}
\text{Var}(D) &= \text{Var}\left(\sum_{j=1}^{N} \hat{a}_j\right) \\
&= <N>\text{Var}(\hat{a}) + \text{Var}(N)[E(\hat{a})]^2 \\
&= <N>(E(w^2.f^2) - [E(w.f)]^2) + \text{Var}(N)[E(w.f)]^2 \\
&= <N>(<w^2.f^2> - <w.f>^2) + \text{Var}(N)<w.f>^2
\end{aligned} \tag{3}$$

The percent coefficient of variation is then given as,

$$CV = \frac{100[\text{Var}(D)]^{\frac{1}{2}}}{D} \tag{4}$$

Each of the terms in Eqs. 2 and 3 can be estimated from knowledge of the dose, particle size, and assay distribution with granule particle size. The only term that is non-trivial is $\text{Var}(N)$. However, $\text{Var}(N)$ can also be evaluated exactly as we show below.

The total number fluctuations in a fixed volume $V$ are given by

$$\langle (\Delta N)^2 \rangle = \langle \sum_{i=1}^{m} (N_i - \bar{N}_i)^2 \rangle = \sum_{i=1}^{m} \langle (\Delta N_i)^2 \rangle + \sum_{i \neq j=1}^{m} \langle (\Delta N_i)(\Delta N_j) \rangle \tag{5}$$

In the rest of the paper, summation over all $m$ species will not be shown explicitly and is understood. In the grand canonical ensemble it can be shown that,[27]

$$\langle (\Delta N_i)^2 \rangle = \left.\frac{\partial N_i}{\partial \beta \mu_i}\right|_{V,T,\mu_{k,k \neq i}} \quad \text{and} \quad \langle (\Delta N_i)(\Delta N_j) \rangle = \left.\frac{\partial N_i}{\partial \beta \mu_j}\right|_{V,T,\mu_{k,k \neq j}} \tag{6}$$

where $\beta = 1/kT$; $k$ is the Boltzmann's constant and $T$ is the temperature in Kelvin. Thus, the total number fluctuations can be obtained from the individual species number fluctuations and cross species number fluctuations given by Eq. 6. For a two-component system it is known that the matrix of the individual species and cross species number fluctuations is the inverse of the matrix formed with the derivative of the chemical potentials.[28-30] It is easy to prove that this result is also applicable to multicomponent systems. Therefore, we have,

$$\begin{pmatrix} \frac{\partial N_1}{\partial \beta \mu_1}\Big|_{V,T,\mu_{k,k\neq 1}} & \cdots & \frac{\partial N_1}{\partial \beta \mu_n}\Big|_{V,T,\mu_{k,k\neq n}} \\ \vdots & \ddots & \vdots \\ \frac{\partial N_n}{\partial \beta \mu_1}\Big|_{V,T,\mu_{k,k\neq 1}} & \cdots & \frac{\partial N_n}{\partial \beta \mu_n}\Big|_{V,T,\mu_{k,k\neq n}} \end{pmatrix} = \begin{pmatrix} \frac{\partial \beta \mu_1}{\partial N_1}\Big|_{V,T,N_{k,k\neq 1}} & \cdots & \frac{\partial \beta \mu_1}{\partial N_n}\Big|_{V,T,N_{k,k\neq n}} \\ \vdots & \ddots & \vdots \\ \frac{\partial \beta \mu_n}{\partial N_1}\Big|_{V,T,N_{k,k\neq 1}} & \cdots & \frac{\partial \beta \mu_n}{\partial N_n}\Big|_{V,T,N_{k,k\neq n}} \end{pmatrix}^{-1} \quad (7)$$

In order to calculate the matrix $\left[\frac{\partial \beta \mu}{\partial N}\right]$ an analytical theory is needed. Fortunately, such a theory is readily available in the literature. For a system of pure component hard spheres, Wertheim[14] and Thiele[15] succeeded in obtaining an exact solution of the P-Y integral equation for the radial distribution function g($r$). For a system consisting of a mixture of hard spheres, Lebowitz[16,17] has obtained an exact solution, which we use here in the calculation for the individual and cross species fluctuations. The expression for the chemical potential is obtained as,[17]

$$\beta \mu_i = \ln \left[ \frac{\rho_i h^3}{(2\pi m_i kT)^{\frac{3}{2}}} \right] + \ln(1-\xi) + \frac{\pi}{6}\beta P R_i^3 + \frac{R_i^2}{(1-\xi)^3}\left\{3Y - 6\xi Y + \frac{9}{2}X^2 - \frac{9}{2}\xi X^2 + 3\xi^2 Y\right\} + \frac{R_i}{(1-\xi)^3}\{3X - 6\xi X + 3\xi^2 X\} \quad (8)$$

and

$$\beta P = \left\{[\Sigma \rho_i][1 + \xi + \xi^2] - \left(\frac{18}{\pi}\right) \sum_{i<j} \eta_i \eta_j (R_i - R_j)^2 [2R_{ij} + XR_i R_j]\right\}(1-\xi)^{-3} \quad (9)$$

where

$R_i$ is the diameter of a particle of the $i$th species, $m_i$ is the mass of a particle of the $i$th species, $\rho_i = N_i/V$ is the number density of the $i$th species, and $h$ is Planck's constant.

$$R_{ij} \equiv (R_i + R_j)/2 \,,\ \eta_i = \frac{1}{6}\pi \rho_i \,,\ \xi = \sum \eta_i R_i^3 \,,\ X = \sum \eta_i R_i^2 \,,\ Y = \sum \eta_i R_i \quad (10)$$

Here $\xi$ is the packing fraction i.e. the volume fraction occupied by granules in the tablet. From differentiation of Eq. 8, the matrix $\left[\frac{\partial \beta \mu}{\partial N}\right]$ can be easily constructed. Then the individual and cross species fluctuations can be calculated exactly through Eq. 7. This allows for the evaluation of $CV$ from Eqs. 3-4. It is also possible to calculate $CV$ directly from the individual and cross species number fluctuations. An example to show this calculation is included in Results-Tableting study section.

**Tableting study**

To evaluate the predictions of the theoretical approach presented herein with experimental data an extensive tablet design study was conducted. To capture the impact of granule particle size on CU, two lots of granules - batch A and B - were manufactured, with batch A being smaller than batch B. The volume mean diameter of batch A granules was determined to be 317 micron, while for batch B granules it was 391 micron. Tablets with varying weight and loading of granules for each of the two granule batches were manufactured. The experimental plan of 9 tablet formulations for each granule batch is outlined in Table 1. A total of 54 independent tablet batches were manufactured to ensure that a reasonably large data set is available (three tablet batches were manufactured for each formulation). Thus, the tablet design study systematically explores the impact of three factors viz., the effect of percentage loading, tablet weight, and granule particle size on tablet CU.

Diphenhydramine HCL was chosen as a model API. The API was granulated in a high shear granulator (KG-5L, Key International, Englishtown, NJ) and coated with Eudragit E100 dissolved in ethyl alcohol in a bottom spray fluid bed processor (MP-1, Niro Aeromatic, Columbia, MD). The target potency for the coated granules was ~30% w/w. The granules were screened, and, for batch A particles larger than 30 mesh or smaller than 100 mesh were discarded. While for batch B, particles larger than 25 mesh or smaller than 120 mesh were discarded. The coated granules were blended with mannitol 60, granular mannitol 2080, microcrystalline cellulose, and crospovidone in a V-blender (Patterson-Kelley, East Stroudsburg, PA) for 30 minutes. Following which magnesium stearate was added and the mixture was further blended for an additional 5 minutes. The granules were loaded at 5.0, 12.5, and 25.0% w/w. The loading of the excipients were as follows: microcrystalline cellulose (10.0% w/w), crospovidone (6.0 % w/w), magnesium stearate (1.5 % w/w). Granular mannitol 2080 and mannitol 60 were loaded at 1:1 ratio to add up all ingredients to 100%. The blend was then compressed on a rotary tablet press (Piccola B10-C, Riva S.A. Argentina) using 4 stations at 40 and 25 rpm press and paddle speeds respectively. Batch size was $\geq$ 3800 tablets for each formulation studied. Tablet samples were collected in ten approximately equally spaced intervals throughout the batch. Three tablets per interval were analyzed for API content using UV/VIS spectrophotometry (Model 8453, Agilent Technologies, Santa Clara, CA) at wavelength of 258 nm. For each tablet formulation three independent batches were manufactured and analyzed for CU. The coated granules were also analyzed for particle size distribution using several screen sizes (25 – 120 US mesh) and assay for each sieve fraction was determined using spectrophotometry. The granule particle size and assay data are shown in Table 2.

**RESULTS**

**Theoretical model**

**Exact solution:**

The most direct evaluation of the *CV* without any approximations can be obtained by writing the variance in the drug content as follows:

$$\text{Var}(D) = \left\langle \left( \sum_i a_i N_i - \sum_i a_i \bar{N}_i \right)^2 \right\rangle$$
$$= \sum_i a_i^2 \langle (\Delta N_i)^2 \rangle + \sum_{i \neq j} a_i a_j \langle (\Delta N_i)(\Delta N_j) \rangle \quad (11)$$

The above method of calculating the *CV* is exact; however, it does not give us any insight into the different granule and tablet properties that collectively impact the *CV*. To understand that relationship between the variables involved we have to obtain an approximate expression for Var(*N*) in Eq. 3.

**Approximate solutions:**

In this section we show how an approximate expression for Var(*N*) can be derived.

**i)** *Single component system*

It is well known that for a pure component, the number fluctuations in a control volume *V* are given exactly as follows:[31]

$$\frac{\langle (\Delta N)^2 \rangle}{N^2} = \frac{\kappa_T}{\beta V} \quad (12)$$

where $\kappa_T = \frac{-1}{V} \frac{\partial V}{\partial P}\Big|_{T,N}$ is the isothermal compressibility. For the evaluation of the isothermal compressibility, we use the equation of state (EOS) for pure component hard sphere systems obtained from the integral equation theory by Wertheim[14] and Thiele[15],

$$\frac{\beta P V}{N} = \frac{1 + \xi + \xi^2}{(1-\xi)^3} \quad (13)$$

This EOS results in the following expression for the isothermal compressibility, when terms up to $\xi^2$ are retained:

$$\frac{\kappa_T}{\beta V} = \frac{1}{N(1 + 8\xi + 21\xi^2)} \quad (14)$$

Plugging Eq. 14 in Eq. 12 we find that the Var(*N*) for a pure component is given as:

$$\langle (\Delta N)^2 \rangle = \frac{N}{(1 + 8\xi + 21\xi^2)} \quad (15)$$

Since both *w* and *f* are fixed for a pure component species, inserting Eq. 15 into Eqs. 2-4 we find that

$$CV = \frac{100}{\sqrt{D}} \left[ \frac{w.f}{1 + 8\xi + 21\xi^2} \right]^{1/2} \quad (16)$$

Equation 16 is exact as it is derived from an analytical theory for single component hard spheres that has been shown to reliably describe fluid behavior up to at least a packing fraction of $\xi \lesssim 0.4$.[17]

**ii) *Multicomponent system***

For a multicomponent system, the term $\frac{\kappa_T}{\beta V}$ no longer represents the total number fluctuations. For example, for a two component system (species 1,2), it can be shown with straightforward thermodynamics that:[28-30]

$$\frac{\langle(\Delta N)^2\rangle}{N^2} = \frac{\kappa_{T,N_1,N_2}}{\beta V} - \frac{N_1 N_2 (\bar{v}_2 - \bar{v}_1)^2}{(\partial \beta \mu_1 / \partial N_2)_{T,P,N_1}} \tag{17}$$

where, $\bar{v}_i$ denotes the partial molar volume of species *i*. The cross-derivative $(\partial \beta \mu_1 / \partial N_2)_{T,P,N_1}$ is non-positive, which makes the second term in Eq. 17 positive.[32] Thus, in this case, the term $\frac{\kappa_T}{\beta V}$ captures only a portion of the total number fluctuations.

For mixtures with >2 components, the expansion of total number fluctuations into contributions from compressibility and other contributions is non-trivial and is not discussed here. However, the two component mixture serves as a good example to highlight the fact that the total number fluctuations cannot be entirely derived from the isothermal compressibility for multicomponent systems. Therefore, unlike Eq. 16, which is exact for a single component system, only an approximate analytical expression for the *CV* can be obtained for multicomponent mixtures. However, in the following we find that in multicomponent systems for small $\xi \lesssim 0.2$, the term $\frac{\kappa_T}{\beta V}$ serves as a good approximation for the total number fluctuations. This approach to estimating the total number fluctuations allows us to understand the various factors that contribute to the *CV* in tablets and blends.

From the EOS in Eq. 9 the isothermal compressibility for a multicomponent system of hard spheres can be obtained (only terms up to $\xi^2$ are retained).[17,18] Therefore,

$$\frac{\langle(\Delta N)^2\rangle}{N^2} \cong \frac{\kappa_T}{\beta V} = \frac{1}{N(1 + 8\xi + 21\xi^2 - 6y_1\xi - 27y_1\xi^2 - 9y_2\xi^2)} \tag{18}$$

Where,

$$y_1 = \sum_{j>i=1}^{m} \Delta_{ij}(R_i + R_j)(R_i R_j)^{-1/2}$$

$$y_2 = \sum_{j>i=1}^{m} \Delta_{ij} \sum_{k=1}^{m} \left(\frac{\xi_k}{\xi}\right)\frac{(R_i R_j)^{1/2}}{R_k}$$

$$\Delta_{ij} = [(\xi_i \xi_j)^{1/2}/\xi][(R_i - R_j)^2/R_i R_j](x_i x_j)^{1/2}$$

$$\xi_i = \frac{1}{6}\pi \rho_i R_i^3$$

and $x_i$ is the mole fraction of the *i*th species.

Note that Eq. 18 is exact as it is derived from an analytical theory for hard sphere mixtures that has been shown to reliably describe fluid behavior up to at least a packing fraction of $\xi \lesssim 0.51$.[18]

Inserting Eq. 18 into Eq. 3, we find that

$$\text{Var}(D) = N \left[ <w^2.f^2> - <w.f>^2 \frac{(8\xi + 21\xi^2 - 6y_1\xi - 27y_1\xi^2 - 9y_2\xi^2)}{(1 + 8\xi + 21\xi^2 - 6y_1\xi - 27y_1\xi^2 - 9y_2\xi^2)} \right] \quad (19)$$

This shows that the $CV$ is given as

$$CV = \underbrace{\frac{100}{\sqrt{D}}}_{1} \underbrace{\left[ \frac{<w^2.f^2>}{<w.f>} \right.}_{2} - \underbrace{<w.f> \frac{(8\xi + 21\xi^2 - 6y_1\xi - 27y_1\xi^2 - 9y_2\xi^2)}{(1 + 8\xi + 21\xi^2 - 6y_1\xi - 27y_1\xi^2 - 9y_2\xi^2)} \right]^{1/2}}_{3} \quad (20)$$

The different grouped terms in the above equation are discussed in further detail in Discussion-Theoretical model section.

**Tableting study**

To show the predictions from the model and its comparison with the experimental data, at first the true density of the coated granules had to be obtained. This was performed by taking ~25 gm of the granules in a volumetric cylinder. To this, a non-interacting solvent (squalene) was added and mixed to displace all the air bubbles. The final volume of the mixture was noted. Taking into account the final volume of the mixture, density of squalene, and weight of granules, the true density of the granules for batch A and B was calculated to be 1.2 gm/cc. We assume that the true density for the different sieve cuts are very close and therefore use a single value for all based on the composite.

Next, the packing fraction, $\xi$, in each tablet formulation had to be estimated. This was done by taking into account the blend density ($\cong$ 0.6 gm/cc, irrespective of granule loading), tablet weight, w/w percent loading of granules, and true density of granules. The calculated values are shown in Table 1.

In Table 3 we show the calculations performed in estimating the various terms in Eq. 20. For each sieve cut, we use the median diameter of the particles for the calculation. In Table 3, the number fraction of granules of each sieve cut were obtained by dividing the weight fraction with the cube of the median diameter and normalizing the ratios to add up to 1.0. This method is based on the assumption that the true density of each sieve fraction is identical. The weight of each bead was obtained by multiplying the volume of each bead with the true density. We obtain the following properties for the granules, batch A: $<w^2.f^2> = $ 2.85E-11 gm$^2$ and $<w.f> = $ 2.95E-06 gm and batch B: $<w^2.f^2> = $ 7.40E-11 gm$^2$ and $<w.f> = $ 4.61E-06 gm. The number of granules of each sieve cut in a tablet, and the associated number densities, $\rho_i$, in a tablet for use in Eqs. 8 and 9 can be calculated from the volume of the granules of each sieve cut in a tablet and the volume of each granule of the sieve cut. From Eq. 18, $y_1$ and $y_2$ were calculated to be 0.1961 and 0.0773 respectively for batch A, and 0.2495 and 0.0991 respectively for batch B.

With the information in Tables 1, 2 and 3, the predicted *CV* from Eqs. 11 and 20 can be easily calculated. The predicted *CV*s from these equations, along with the experimentally measured tablet weight corrected *CV* (n=3 batches for each formulation) are shown in Table 4.

In these calculations, to obtain the individual number fluctuations, the matrix $\left[\frac{\partial \beta \mu}{\partial N}\right]$ was calculated by numerical differentiation of Eq. 8. The inverse matrix, viz., $\left[\frac{\partial N}{\partial \beta \mu}\right]$ was obtained using the MINVERSE function in MS Excel. Then, the exact *CV* can be easily obtained through Eq. 11.

**Note:** Each sieve fraction shown in Table 2 is considered as a species for the calculation with the theoretical model. This can be done because each sieve cut is approximated by the median diameter for its particles and they have a given assay fraction (*f*). Thus, we consider that are eight species of granules (*m*=8) within each granule batch A or B. However, the splitting of any granule batch into number of species is arbitrary (depending upon the sieves used) and the model can handle any level of approximation ($0 < m < \infty$). The sieves shown in Table 2 were chosen because they allow for maximum resolution in the particle size range. The explicit calculations are shown in Table 3.

**Comparison to literature**

The theoretical model requires a large set of information for making predictions, such as for the granules (true density, particle size distribution, assay distribution), blend (bulk density, percentage granule load), and tablet weight. Unfortunately, at present, there is no study in the literature that reports all this data along with the experimental tablet *CV*. The only study that comes close to providing all required information is by Egermann et al.[8,9] In this study, Egermann and coworkers used sucrose as a model drug in a mixture of sucrose/microcrystalline cellulose/talc and obtained the *CV* of sucrose in 50 mg and 200 mg tablets. The following ratios (w/w) of sucrose and microcrystalline cellulose/talc in the blend were studied: 10:90, 30:70, 50:50, and 80:20. The authors only reported the volume mean particle size of the ingredients ($d_{50}$) as 504 micron and 60 micron for sucrose and microcrystalline cellulose/talc, respectively. Using this data and the reported values for other quantities, predictions for the *CV* from the theoretical model are shown in Table 5, along with predictions from Egermann's model and experimental *CV*. Note that since only the $d_{50}$ is available for the sucrose, it was approximated as a single species in the theoretical model and *CV* predictions were made using Eq. 16. Results from Table 5 show that the predictions of the present model reasonably agree with the predictions of the Egermann model.

**DISCUSSION**

**Theoretical model**

Equation 20 allows us to clearly understand the contributions from various factors to the final content uniformity. The first ratio shows that the *CV* varies inversely with the square root of the dose if all other factors are fixed (granule particle size and assay distribution, granule loading). The second ratio shows how the granule characteristics with respect to particle size distribution and assay distribution with particle size impact the *CV*. The third term shows how the percent loading of granules (on a vol/vol basis) affects the *CV*, although, this effect is coupled with the assay distribution of the granules through the term <w.f> and the particle size distribution through the terms $y_1$ and $y_2$. It can be seen that the $CV$ is a complex function of the granule particle size and assay distribution, tablet volume, and granule loading. Therefore, it is not possible to generate a simplified nomograph e.g. for the limiting granule particle size distribution for passing the USP CU criteria as is available in the ideal mixing approximation from references 5-7. In this case, the limiting granule characteristics have to be obtained for each tablet formulation individually from the theory.

Further insight from Eq. 20 can be obtained as follows. If the dose is fixed, but the tablet size is increased such that $\xi \to 0$ (this will be the situation in low granule loading or low dose tablets), Eq. 20 evolves as:

$$CV_{\xi \to 0} = \frac{100}{\sqrt{D}} \left[ \frac{<w^2.f^2>}{<w.f>} \right]^{1/2} \tag{21}$$

Equation 21 shows that for a given dose the $CV$ is determined by the ratio $\lambda = <w^2.f^2>/<w.f>$. Thus, clearly, $\lambda$ characterizes the "*granule quality*". If $\lambda$ is larger the $CV$ is larger and vice versa. To ensure a low $CV$ we must have as small $\lambda$ as possible. From Table 3 we find that $\lambda$=9.66E-06 and 1.60 E-05 gm for granule batch A and B, respectively.

If the particle size and its associated assay fraction are truly independent, one can write $\lambda$ as:

$$\lambda = \frac{(Var(w) + <w>^2)(Var(f) + <f>^2)}{<w>.<f>} \tag{22}$$

Where $Var(w) = <w^2> - <w>^2$ and $Var(f) = <f^2> - <f>^2$ denote the variance in the particle size distribution and assay with particle size distribution. Equation 22 shows that both the particle size distribution and assay with particle size distribution equally influence the granule quality and hence the $CV$ of tablets. Thus it is not merely sufficient to have a narrow particle size distribution; attention should also be paid to the assay distribution. <u>Note</u>: If the particle size and its associated assay fraction are not independent, or, their degree of independence is not known, Eq. 22 should not be used for calculation of $\lambda$. The exact definition of $\lambda$ should be used instead.

A special case of Eq. 21 is when *f* is fixed (for granules) or *f*=1 (as in pure API particles). Then, the equation reduces to:

$$CV_{\xi \to 0, f=fixed} = \frac{100}{\sqrt{D}} \left[ \frac{f.<w^2>}{<w>} \right]^{1/2} \tag{23}$$

Equation 23 is identical to one recently derived by Huang and Ku.[7] This equation was shown to reliably predict the $CV$ in low dose formulations with pure API particles.

Another observation from the present study is that it is not always true that removing only larger particles will reduce the $CV$. For example, consider two cases for illustration i) if the particle assay is increasing with size and ii) when particle assay is decreasing with size. Representative plots for the product *w.f* of each particle with the number fraction in these two granule systems are shown in Figure 2. Both granule systems have a fixed log-normal particle size distribution. The particle size is increasing on the x-axis from left to right for granules i while is it is decreasing in the same direction for granules ii. For both granules, it is found that $\lambda$ can be decreased by removing particles of large values for *w.f*, which for granules ii is actually the smaller size particles. In our opinion, for any granule system removing particles of larger values for *w.f* should reduce $\lambda$. However, this needs to be evaluated further and could be verified on a case by case basis.

**Tableting study**

In Table 4, we find that the $CV$ calculated through Eq. 20 is close to the $CV$ obtained from the exact expression through Eq. 11. This closeness provides further proof that Eq. 20 is able to capture majority of the number fluctuations in typical blend or tablet formulations with relatively low packing fraction ($\xi \lesssim 0.2$). It can be shown that with increasing $\xi$ the fraction of the number fluctuations captured by the term $\frac{\kappa_T}{\beta V}$ decreases. For example, in granule batch A it is found that the percentage of the total number fluctuations captured by the term $\frac{\kappa_T}{\beta V}$ at $\xi = 0.026, 0.065, 0.13$ and $0.2$ are 92, 82, 67, and 50% respectively. Similar results are obtained for granule batch B tablet formulations. Therefore, at larger $\xi$, Eq. 20 becomes a less accurate approximation of Eq. 11. This is also apparent in Table 4 where the difference in the $CV$ calculated through Eq. 11 and Eq. 20 increases with increasing granule loading i.e. $\xi$, for a fixed tablet weight.

It should also be noted that during regular manufacturing a distribution in the measured $CV$ exists due to the finite number of tablets that are analyzed in each batch, and also because there is batch to batch variability. The theoretical calculation of the $CV$ through Eq. 11 gives the exact mean $CV$ expected for an infinite number of tablets or batches assuming random mixing. The measured mean $CV$ from production batches will always be higher than the predicted mean $CV$ by the model. In addition, the model is not applicable in situations where the particle size of the API or granules is changing due to either agglomeration or de-agglomeration during processing operations.[33,34]

During manufacturing the $CV$ may increase due to segregation in blending operations, dispensing, tableting operation, and other sources. Therefore, we have,

$$CV_{obs}^2 = CV_{calc}^2 + CV_{Tableting}^2 + CV_{Blend\ dispensing}^2 + CV_{Error}^2$$

Or

$$CV_{obs}^2 = CV_{calc}^2 + CV_{All\ errors}^2$$

In the last column of Table 4 we calculate $CV_{All\ errors}^2$ as the difference between the exact $CV$ (through Eq. 11) and the experimental mean $CV$. We find that $CV_{All\ errors}^2$ is roughly same for all the tablet formulations of a given weight. This is interesting as it suggests that the granule loading does not have a large impact on the degree of segregation. Two exceptions to this observation are the results for the 100 mg and 250 mg tablets made with granule batch B at 5% loading. In those cases, $CV_{All\ errors}^2 \cong 0.8$ and $0.9$ respectively from Table 4. However, these results can be understood if we keep in mind that the experimental data was obtained for three independent batches and it is possible that the mean $CV$ may change slightly if more batches are manufactured. For example, for the 5% loading in the 100 mg tablet since the observed mean % $CV$ is high (10.5), even a small rise in the mean $CV$ (for e.g. to 11.1) can result in $CV^2_{All\ errors} \cong 13.8$, which is in agreement with the rest of the data. Given that the $CV$ values obtained for the three individual tablet batches of this formulation are 11.2, 11.0, and 9.4, this is a reasonable possibility.

With both granule batches it is found that with increasing tablet weight $CV^2_{All\ errors}$ significantly decreases. This suggests that the segregation arising from tableting operation can be reduced by making a larger tablet for a given dose. Although at fixed dose and increasing tablet size, the granule loading will decrease which will increase $CV_{calc}$. The increase in $CV_{calc}$ may or may not offset the gain obtained from reduction in $CV^2_{All\ errors}$ and would need to be evaluated on a case by case basis.

The larger values for $CV^2_{All\ errors}$ for the 100 mg tablet are not unexpected considering that filling a smaller tablet die is likely to be more difficult and involve more movement for the blend which can cause more segregation in the placement of granules in tablets. The results from Table 4 show that one can calculate $CV^2_{All\ errors}$ and that may allow us to predict the experimental $CV$ (in combination with the theoretical model) for other products, or, at least other formulations with similar characteristics.

It should be noted that the theoretical model has no fitting parameters, but is a direct calculation. The $CV^2_{All\ errors}$ represents all the contributions to the CU due to inefficient processing operations.

**Comparison to literature**

From Table 5 it is found that the predictions from the present model reasonably agree with the predictions from the model by Egermann.[8,9] Although, for some formulations, predictions from both models are significantly lower than the experimental tablet $CV$ (such as the 200 mg tablet with 10% w/w sucrose). This large difference in the $CV$s could be due to the large difference in particle size of the sucrose (504 micron) and microcrystalline cellulose/talc (60 micron) which is known to cause segregation problems during blend processing. In addition, the sucrose level is low, which is likely to worsen the effect of segregation. Thus, comparison to literature models and data shows that the current model is also consistent with them. However, the current model derives the factors affecting the $CV$ from first principles and takes into account several features such as particle size distribution, assay distribution with particle size, among others, which is not available elsewhere. Note that in Table 5, the predicted $CV$ from the theoretical model is always lower than the experimental $CV$ because the blend will always have some degree of non-randomness.

**CONCLUSIONS**

A generic and exact model for predicting content uniformity of granules in blend and tablet formulations is presented. The model exploits the fundamental theory behind particle number fluctuations in statistical mechanics to allow for the most accurate prediction of the mean $CV$ under the assumption of random mixing. The predictions of the model serve as a theoretical lower limit for the mean $CV$ (for infinite batches or tablets) that can be expected during manufacturing. The measured mean $CV$ from production batches will be higher than the theoretical lower limit. A distinguishing feature of the present model is that it takes into account the excluded volume of the granules in the formulation, and thus, allows us to study the effect of granule loading on the $CV$. Earlier models have mostly been based on the assumption of ideal mixing which ignores the excluded volume, and consequently, are incapable of incorporating the effect of granule loading. Our model also applies to formulations made with API particles as a special case. We also manufactured several tablet formulations with two lot of granules made from a model API to test the predictions from the theoretical model. It was found that the model is able to predict the experimental $CV$ with good success and captures the effect of all elements (such as granule particle size, assay distribution with particle size, percentage loading, and tablet weight) that impact the CU as illustrated in the experimental data. An exact method for calculating the theoretical $CV$ was presented along with analytical expressions which are an approximate solution. The approximate solutions give insight into the factors that contribute to the $CV$. It is shown that the $CV$ varies inversely with the square root of the dose if all other factors are fixed (granule particle size and assay distribution, granule loading). We showed how the percent loading of granules (on a vol/vol basis) affects the $CV$, and, this effect is coupled with the characteristics of the granules. It was established that the $CV$ is determined by the ratio $\lambda = <w^2.f^2>/<w.f>$ which characterizes the "*granule quality*". If $\lambda$ is larger, the $CV$ is larger, and vice versa. To ensure a low $CV$ we must have as small $\lambda$ as possible. This ratio clearly shows the relation between granule properties and observed $CV$. It was argued that it is not always necessary to remove particles with larger size to reduce $CV$, but, instead remove particles with larger values for *w.f* which could be the smaller particles if the assay is decreasing with increasing particle size. Predictions from the current model also agreed reasonably with other models and literature data for $CV$.

The theoretical method presented could be used to first obtain an estimate for $CV^2_{All\ errors}$ in the blend processing and tableting operations. Using the theoretical model and estimate for $CV^2_{All\ errors}$ it will be possible to then predict the experimental $CV$ for products with similar characteristics (e.g. tablet weight, excipient size distribution). In summary, our theoretical model can help in guiding the formulations scientist for designing robust processes to meet the requirements for content uniformity.

**ACKNOWLEDGEMENT**
S.S.R would like to thank Tim Hundertmark, Vikas Agarwal, Jim Netz, Terry Anderson, and Brian Ullmann for providing computer resources and support in execution of the model. S.S.R would also like to thank Nicholas Pick for conducting the API particle size and blend density measurements.

**GLOSSARY**

| | |
|---|---|
| $w_i$ | weight of each granule of type $i$ |
| $f_i$ | fractional (w/w) assay of API in granule of type $i$ |
| $m$ | total number of granule types (i.e. sizes) |
| $\mu_i$ | chemical potential of granule of type $i$ |
| $R_i$ | diameter of granule of type $i$ |
| $V$ | volume corresponding to one tablet in the blend |
| $N_i$ | number of granules of type $i$ in the tablet |
| $\bar{N}_i$ | average number of granules of type $i$ in the tablet |
| $N$ | total number of granules in the tablet |
| $a_i$ | weight of the API in each granule of type $i$ |
| $\hat{a}_j$ | weight of the API in granule $j$ |
| $\beta$ | $1/kT$ |
| $k$ | Boltzmann's constant |
| $T$ | temperature in Kelvin |
| $P$ | pressure |
| $m_i$ | mass of a granule of type $i$ |
| $\rho_i$ | number density of granule of type $i$ |
| $h$ | Planck's constant |
| $\xi$ | packing fraction, i.e. volume fraction occupied by all granules in the tablet |
| $D$ | dose |
| $\langle(\Delta N_i)^2\rangle$ | mean square number fluctuations in granule of type $i$ |
| $\langle(\Delta N_i)(\Delta N_j)\rangle$ | mean product of number fluctuations in granule of types $i$ and $j$ |
| $\langle(\Delta N)^2\rangle$ | total number fluctuations in volume $V$ |
| $\kappa_T$ | isothermal compressibility |
| $\bar{v}_i$ | partial molar volume of granule type $i$ |
| $x_i$ | mole fraction of granule of type $i$ |
| $\xi_i$ | packing fraction of granule of type $i$ |
| $CV$ | percent coefficient of variation in the dose |
| $CV_{obs}$ | percent coefficient of variation in the dose observed from experiments |
| $CV_{calc}$ | percent coefficient of variation in the dose calculated from theory |
| $CV_{All\ errors}$ | contribution to percent coefficient of variation in the dose from segregation during manufacturing operations |
| $\lambda$ | parameter that characterizes "granule quality" |

**Figure 1**. Scheme illustrating the approach of considering each blend volume equivalent to a tablet as an open system within the entire blend. The system is then modeled based on it's analogy to thermodynamics of fluid mixtures.

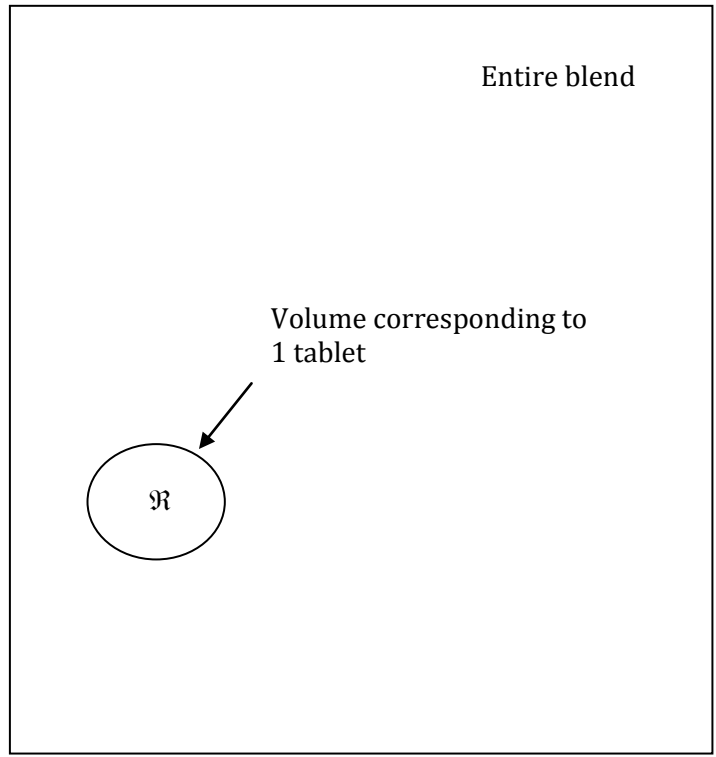

**Figure 2.** An illustrative result for the possible distribution of the product *w.f* in granules, i) assay increasing with particle size, and, ii) assay decreasing with particle size. The particle size distribution is same for both granules. The particle size is increasing on the x-axis from left to right for granules i while is it is decreasing in the same direction for granules ii. For both granules, $\lambda$ can be decreased by removing particles of large values for *w.f*

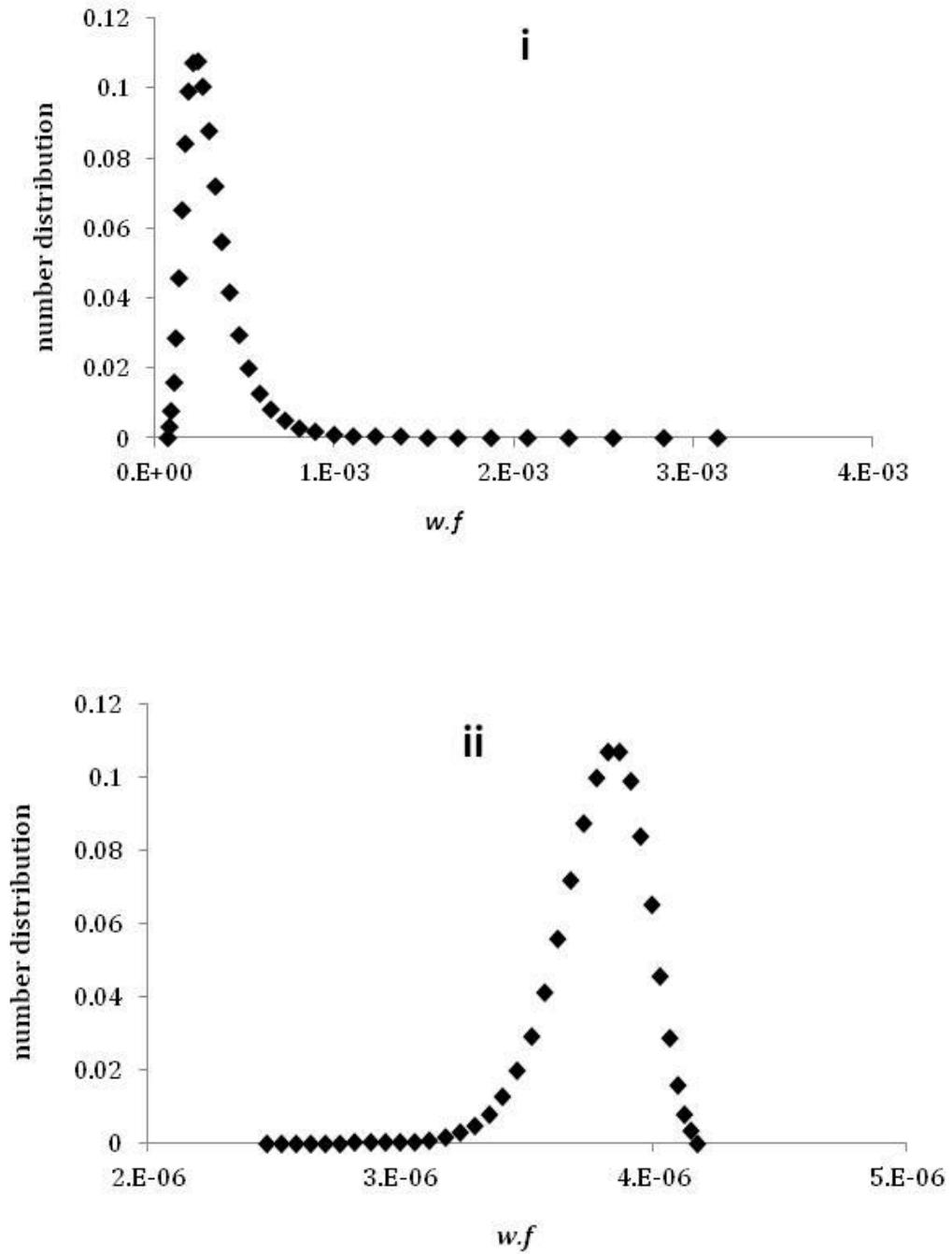

**Table 1.** Experimental plan for tablet manufacture

| Formulation number | Particle loading (%w/w) | Dose (mg) **Batch A granules** | Dose (mg) **Batch B granules** | Tablet weight (mg) and tooling size (inch) | $\xi$ |
|---|---|---|---|---|---|
| 1 | 5.0 | 1.49 | 1.42 | 100 (1/4) | 0.026 |
| 2 | 12.5 | 3.72 | 3.54 | | 0.065 |
| 3 | 25.0 | 7.44 | 7.08 | | 0.13 |
| 4 | 5.0 | 3.72 | 3.54 | 250 (5/16) | 0.026 |
| 5 | 12.5 | 9.30 | 8.84 | | 0.065 |
| 6 | 25.0 | 18.60 | 17.69 | | 0.13 |
| 7 | 5.0 | 7.44 | 7.08 | 500 (1/2) | 0.026 |
| 8 | 12.5 | 18.60 | 17.69 | | 0.065 |
| 9 | 25.0 | 37.20 | 35.38 | | 0.13 |

**Table 2**. Particle size distribution and assay for each sieve fraction for the coated granules

**Batch A**

| Mesh size | Diameter of granules (cm) | Percent retained (w/w) | Assay fraction of each cut |
|---|---|---|---|
| 30 | 0.0595 | 0.0 | NA |
| 35 | 0.05 | 11.4 | 0.3317 |
| 40 | 0.04 | 9.7 | 0.3285 |
| 45 | 0.0354 | 13.1 | 0.3132 |
| 50 | 0.0297 | 14.9 | 0.3081 |
| 60 | 0.025 | 14.9 | 0.3014 |
| 70 | 0.021 | 14.9 | 0.2959 |
| 80 | 0.0177 | 12.4 | 0.2929 |
| 100 | 0.0149 | 8.7 | 0.2871 |
| -30/+100 composite | NA | 100 | 0.2976 |

**Batch B**

| Mesh size | Diameter of granules (cm) | Percent retained (w/w) | Assay fraction of each cut |
|---|---|---|---|
| 25 | 0.0707 | 0.0 | NA |
| 30 | 0.0595 | 9.4 | 0.3023 |
| 35 | 0.05 | 15.4 | 0.3024 |
| 40 | 0.04 | 16.1 | 0.3011 |
| 45 | 0.0354 | 18.5 | 0.2969 |
| 50 | 0.0297 | 15.6 | 0.2939 |
| 60 | 0.025 | 1.9 | 0.2882 |
| 80 | 0.0177 | 20.8 | 0.2848 |
| 120 | 0.0125 | 2.3 | 0.2610 |
| -25/+120 composite | NA | 100 | 0.2826 |

**Table 3**. Calculation of granule characteristics

**Batch A**

| Sieve cut US mesh | Median Diameter (cm) | % w/w retained | Number fraction, p | Assay fraction, f | Volume of each bead, v (cm³) | Weight of each bead, w (gm) | API in each bead, a (gm) | =p.w.f (gm) | =p.w².f² (gm²) |
|---|---|---|---|---|---|---|---|---|---|
| -30/+35 | 0.05475 | 11.4 | 0.01066 | 0.3317 | 8.59E-05 | 1.03E-04 | 3.42E-05 | 3.64E-07 | 1.24E-11 |
| -35/+40 | 0.045 | 9.7 | 0.01628 | 0.3285 | 4.77E-05 | 5.72E-05 | 1.88E-05 | 3.06E-07 | 5.75E-12 |
| -40/+45 | 0.0377 | 13.1 | 0.03733 | 0.3132 | 2.80E-05 | 3.36E-05 | 1.05E-05 | 3.93E-07 | 4.13E-12 |
| -45/+50 | 0.03255 | 14.9 | 0.06622 | 0.3081 | 1.80E-05 | 2.16E-05 | 6.65E-06 | 4.41E-07 | 2.93E-12 |
| -50/+60 | 0.02735 | 14.9 | 0.11162 | 0.3014 | 1.07E-05 | 1.28E-05 | 3.86E-06 | 4.31E-07 | 1.66E-12 |
| -60/+70 | 0.023 | 14.9 | 0.18769 | 0.2959 | 6.37E-06 | 7.64E-06 | 2.26E-06 | 4.24E-07 | 9.59E-13 |
| -70/+80 | 0.01935 | 12.4 | 0.26228 | 0.2929 | 3.79E-06 | 4.55E-06 | 1.33E-06 | 3.50E-07 | 4.66E-13 |
| -80/+100 | 0.0163 | 8.7 | 0.30792 | 0.2871 | 2.27E-06 | 2.72E-06 | 7.81E-07 | 2.40E-07 | 1.88E-13 |
| | | | | | | | Sum= | **2.95E-06** | **2.85E-11** |

**Batch B**

| Sieve cut US mesh | Median Diameter (cm) | % w/w retained | Number fraction, p | Assay fraction, f | Volume of each bead, v (cm³) | Weight of each bead, w (gm) | API in each bead, a (gm) | =p.w.f (gm) | =p.w².f² (gm²) |
|---|---|---|---|---|---|---|---|---|---|
| -25/+30 | 0.0651 | 9.4 | 0.00847 | 0.3023 | 1.44E-04 | 1.73E-04 | 5.22E-05 | 4.42E-07 | 2.31E-11 |
| -30/+35 | 0.05475 | 15.4 | 0.02337 | 0.3024 | 8.59E-05 | 1.03E-04 | 3.12E-05 | 7.28E-07 | 2.27E-11 |
| -35/+40 | 0.045 | 16.1 | 0.04404 | 0.3011 | 4.77E-05 | 5.72E-05 | 1.72E-05 | 7.59E-07 | 1.31E-11 |
| -40/+45 | 0.0377 | 18.5 | 0.08579 | 0.2969 | 2.80E-05 | 3.36E-05 | 9.98E-06 | 8.56E-07 | 8.54E-12 |
| -45/+50 | 0.03255 | 15.6 | 0.11268 | 0.2939 | 1.80E-05 | 2.16E-05 | 6.35E-06 | 7.15E-07 | 4.54E-12 |
| -50/+60 | 0.02735 | 1.9 | 0.02359 | 0.2882 | 1.07E-05 | 1.28E-05 | 3.70E-06 | 8.73E-08 | 3.23E-13 |
| -60/+80 | 0.02135 | 20.8 | 0.53240 | 0.2848 | 5.09E-06 | 6.11E-06 | 1.74E-06 | 9.26E-07 | 1.61E-12 |
| -80/+120 | 0.0151 | 2.3 | 0.16967 | 0.2610 | 1.80E-06 | 2.16E-06 | 5.64E-07 | 9.57E-08 | 5.39E-14 |
| | | | | | | | Sum= | **4.61E-06** | **7.40E-11** |

**Table 4**. Predicted % *CV*s obtained from the theoretical model along with the observed tablet weight corrected % *CV* (shown is the mean and for individual batches, n=3)

### Batch A granules

| Particle loading (%w/w) | Tablet weight (mg) | $CV$ Eq. 11 | $CV$ Eq. 20 | $CV$ Observed (tablet weight corrected) | $CV^2_{All\ errors}$ |
|---|---|---|---|---|---|
| 5.0  | 100 | 7.66 | 7.87 | 8.3 (8.9, 7.0, 9.0) | 10.2 |
| 12.5 |     | 4.39 | 4.83 | 6.1 (6.8, 6.4, 5.2) | 17.9 |
| 25.0 |     | 2.53 | 3.30 | 4.7 (4.0, 5.5, 4.7) | 15.7 |
| 5.0  | 250 | 4.85 | 4.98 | 4.7 (4.2, 5.7, 4.3) | NA |
| 12.5 |     | 2.77 | 3.06 | 3.7 (3.7, 3.5, 4.0) | 6.0 |
| 25.0 |     | 1.60 | 2.09 | 2.2 (2.2, 1.8, 2.5) | 2.3 |
| 5.0  | 500 | 3.43 | 3.52 | 3.6 (3.1, 4.0, 3.8) | 1.2 |
| 12.5 |     | 1.96 | 2.16 | 2.6 (2.5, 2.9, 2.4) | 2.9 |
| 25.0 |     | 1.13 | 1.48 | 1.9 (1.8, 2.0, 2.0) | 2.3 |

### Batch B granules

| Particle loading (%w/w) | Tablet weight (mg) | $CV$ Eq. 11 | $CV$ Eq. 20 | $CV$ Observed (tablet weight corrected) | $CV^2_{All\ errors}$ |
|---|---|---|---|---|---|
| 5.0  | 100 | 10.46 | 10.42 | 10.5 (11.2, 11.0, 9.4) | 0.8 |
| 12.5 |     | 6.04  | 6.42  | 7.2 (8.3, 7.9, 5.5)    | 15.4 |
| 25.0 |     | 3.42  | 4.40  | 5.0 (4.6, 5.3, 5.2)    | 13.3 |
| 5.0  | 250 | 6.63  | 6.59  | 6.7 (6.8, 6.7, 6.7)    | 0.9 |
| 12.5 |     | 3.82  | 4.06  | 4.7 (4.8, 4.8, 4.5)    | 7.5 |
| 25.0 |     | 2.16  | 2.78  | 2.9 (2.6, 3.0, 3.2)    | 3.7 |
| 5.0  | 500 | 4.69  | 4.66  | 4.3 (4.0, 4.9, 4.1)    | NA |
| 12.5 |     | 2.70  | 2.87  | 3.0 (3.1, 2.7, 3.3)    | 1.7 |
| 25.0 |     | 1.53  | 1.97  | 2.2 (1.9, 2.4, 2.4)    | 2.5 |

**Table 5**. Predicted mean % *CV* for sucrose in tablets made from sucrose/microcystalline cellulose/talc mixtures along with the observed experimental % *CV* (n=2)

### 50 mg tablets

| Sucrose (%w/w) | Mean %CV predicted by Eq. 16 | Mean %CV from Egermann's theory[8,9] | Experimentally measured %CV |
|---|---|---|---|
| 10 | 13.1 | 14.4 | NA |
| 30 | 6.2 | 8.2 | NA |
| 50 | 3.6 | 4.0 | 3.9,4.1 |
| 80 | 1.7 | 2.2 | 1.8,2.2 |

### 200 mg tablets

| Sucrose (%w/w) | Mean %CV predicted by Eq. 16 | Mean %CV from Egermann's theory[8,9] | Experimentally measured %CV |
|---|---|---|---|
| 10 | 6.6 | 7.1 | 9.7,10.1 |
| 30 | 3.1 | 4.0 | 4.7,4.6 |
| 50 | 1.8 | 2.0 | 2.2,2.1 |
| 80 | 0.9 | 1.1 | 2.0,2.6 |